\documentclass[%
 reprint,
superscriptaddress,
 amsmath,amssymb,
 prd,
]{revtex4-2}
\usepackage[T1]{fontenc}
\usepackage{times}%
\usepackage{macros}

\usepackage{float}
\usepackage{placeins}

\usepackage{txfonts}
\DeclareSymbolFont{cmletters}{OML}{cmm}{m}{it}
\DeclareMathSymbol{v}{\mathalpha}{cmletters}{"76} 
\usepackage{hyperref}
\usepackage{booktabs}
\usepackage{amssymb}
\usepackage{amsmath}
\usepackage[dvipsnames]{xcolor}
\usepackage{xspace}

\newcommand{\chandra}{{\it Chandra}\xspace}
\newcommand{\mdot}{\ensuremath{\dot M}\xspace}
\newcommand{\edot}{\ensuremath{\dot E}\xspace}
\newcommand{\mdotin}{\ensuremath{\dot M_{\rm in}}\xspace}
\newcommand{\mdotout}{\ensuremath{\dot M_{\rm out}}\xspace}

\newcommand{\mdotbh}{\ensuremath{\dot{M}_{\rm BH}}\xspace}
\newcommand{\avmdotbh}{\ensuremath{\langle \dot{M}_{\rm BH}} \rangle \xspace}
\newcommand{\avphibh}{\ensuremath{\langle \phi_{\rm BH} \rangle} \xspace}
\newcommand{\aveta}{\ensuremath{\langle \eta \rangle} \xspace}

\newcommand{\mdotb}{\ensuremath{\dot M_{\rm B}}\xspace}
\newcommand{\rh}{\ensuremath{r_{\rm H}}\xspace}
\newcommand{\rg}{\ensuremath{r_{\rm g}}\xspace}
\newcommand{\tg}{\ensuremath{t_{\rm g}}\xspace}
\newcommand{\tb}{\ensuremath{t_{\rm B}}\xspace}
\newcommand{\rw}{\ensuremath{r_{\rm w}}\xspace}

\newcommand{\rb}{\ensuremath{r_{\rm B}}\xspace}
\newcommand{\rcirc}{\ensuremath{r_{\rm c}}\xspace}
\newcommand{\BH}{{{\rm BH}}\xspace}

\newcommand{\MBH}{{M_{\rm BH}}\xspace}
\newcommand{\phibh}{{\phi_{\rm BH}}\xspace}

\newcommand{\der}{{\rm d}}

\usepackage[normalem]{ulem}

\usepackage{hyperref}

\usepackage{orcidlink}
\usepackage{graphicx}

\usepackage[sort&compress]{natbib}
\defcitealias{Lalakos2024}{L24}

\begin{document}

\title{Universal Radial Scaling of Large-Scale Black Hole Accretion for Magnetically Arrested And Rocking Accretion Disks}

\author{Aretaios Lalakos \orcidlink{0000-0002-6883-6520}}
\email{lalakos@caltech.edu}
\affiliation{TAPIR, Mailcode 350-17, California Institute of Technology, Pasadena, CA 91125, USA}
\affiliation{Walter Burke Institute for Theoretical Physics, California Institute of Technology, Pasadena, CA 91125, USA}
\affiliation{Canadian Institute for Theoretical Astrophysics, 60 St. George Street, Toronto, ON M5S 3H8, Canada}
\author{Alexander Tchekhovskoy \orcidlink{0000-0002-9182-2047}}
\affiliation{Center for Interdisciplinary Exploration \& Research in Astrophysics (CIERA), Physics \& Astronomy, Northwestern University, Evanston, IL 60201, USA}
\affiliation{NSF-Simons AI Institute for the Sky (SkAI), 172 E. Chestnut St., Chicago, IL 60611, USA}
\affiliation{Department of Engineering Sciences and Applied Mathematics, Northwestern University, Evanston, IL 60208, USA}
\author{Elias R. Most \orcidlink{0000-0002-7301-3908}}
\affiliation{TAPIR, Mailcode 350-17, California Institute of Technology, Pasadena, CA 91125, USA}
\affiliation{Walter Burke Institute for Theoretical Physics, California Institute of Technology, Pasadena, CA 91125, USA}
\author{Bart Ripperda \orcidlink{0000-0002-0491-1210}}
\affiliation{Canadian Institute for Theoretical Astrophysics, 60 St. George Street, Toronto, ON M5S 3H8, Canada}
\affiliation{Department of Physics, University of Toronto, 60 St. George Street, Toronto, ON M5S 1A7, Canada}
\affiliation{David A. Dunlap Department of Astronomy, University of Toronto, 50 St. George Street, Toronto, ON M5S 3H4, Canada}
\affiliation{Perimeter Institute for Theoretical Physics, 31 Caroline St. North, Waterloo, ON N2L 2Y5, Canada}
\author{Koushik Chatterjee \orcidlink{0000-0002-2825-3590}}
\affiliation{Institute for Research in Electronics and Applied Physics, University of Maryland, College Park, MD 20742, USA}
\author{Matthew Liska \orcidlink{0000-0003-4475-9345}}
\affiliation{Center for Relativistic Astrophysics, Georgia Institute of Technology, Howey Physics Bldg, 837 State St NW, Atlanta, GA 30332, USA}


\begin{abstract}
Accretion onto supermassive black holes (BHs) can launch relativistic outflows and jets that inject energy and momentum into their surroundings. Understanding how such feedback shapes large-scale accretion is key to bridging observations from galactic scales (e.g., the Bondi radius, $r_{\rm B}$) down to event horizon scales ($r_{\rm g}$), spanning $5{-}6$ orders of magnitude. To address this challenge directly, we treat the spatial scale separation as a free parameter, varying it across $2{-}4$ orders of magnitude. We perform a suite of the longest contiguous 3D general relativistic magnetohydrodynamic (GRMHD) simulations to date ($t \lesssim 4\times10^6\rg/c$), modeling Bondi-like accretion of rotating, non-relativistic gas with weak vertical magnetic fields onto a rapidly spinning BH, achieving inflow equilibrium out to $r \gtrsim 10^3\rg$.
We find that, regardless of scale separation or ambient gas rotation, all simulations reach a magnetically arrested disk (MAD) state in which the BH becomes magnetically saturated. In this state, the mass inflow rate follows a universal radial scaling relative to the Bondi rate: $\dot{M}_{\rm in}(r)/\dot{M}_{\rm B} \sim (r/r_{\rm B})^s$ with $s = 0.66 \pm 0.03$.
The MAD state self-regulates through jets, outflows, and magnetic flux eruptions that can ultimately disrupt coherent angular momentum inflow, giving rise to a rocking accretion disk (RAD) state. This RAD state features chaotically oriented inflows, weak intermittent jets, and a steeper inflow slope of $s = 0.87 \pm 0.05$, along with significantly weaker outflows.
For rapidly spinning BHs, the MAD and RAD BH accretion rates become comparable at typical scale separations, $\rb/\rg \gtrsim 10^5$.
The weaker outflows in the RAD state allow large-scale inflows to resume, eventually restoring the MAD state and enabling a repeating MAD–RAD cycle.
We find that the MAD-RAD timescales roughly lasts tens of Bondi timescales,
$\tb \sim 0.2\,\text{Myr} \times (\rb/10^{5}\rg)^{3/2} \times (M_{\rm BH}/10^9M_\odot)$, where $M_{\rm BH}$ is the BH mass,
potentially setting the duty cycle of jetted active galactic nucleus (AGN) outbursts, like M87*.
\end{abstract}

\maketitle

\section{Introduction} \label{sec:intro} 
Supermassive black holes (SMBHs) feed on the surrounding gas, convert accreted material into energy, and power active galactic nuclei (AGN).
Most AGN accrete at sub-Eddington rates and are classified as low-luminosity AGN (LLAGN). At low accretion rates, BHs are thought to primarily accrete through radiatively inefficient accretion flows (RIAF) \cite{nar94}, which can power relativistic, collimated outflows, or jets \cite{yuan2014hot}.
Magnetized gas accretion onto the BH can lead to jet formation, whose feedback modifies the gas dynamics on the scales much larger than the BH event horizon \cite{ssl07,fabian2012, morganti2017many}, of size $\rg=GM_{\rm BH}/c^2$, where $M_{\rm BH}$ is the BH mass, $G$ is the gravitational constant, and $c$ is the speed of light. 

Jets can extract BH spin energy, via large-scale magnetic fields threading the event horizon \cite{blanford1977}. Slower, broader, and less collimated outflows can also tap into the rotational energy of the accretion disk, via magneto-centrifugal forces \cite{bp82, blandford2019relativistic}. Jets emitting in radio wavelengths can reach immense distances, up to several megaparsecs \cite{2022A&A...660A...2Ow, oei2024black}. As  jets propagate through the interstellar medium (ISM) and the intracluster medium (ICM), they displace gas and inflate X-ray cavities \cite{mcnamara2007heating, mcnamara2012mechanical}. This feedback process can regulate cooling flows through shocks, acoustic waves, and turbulent heating, and play a crucial role in shaping galaxy evolution \cite{2016MNRAS.458.2902Z, martizzi2019simulations, 2020ApJ...889L...1L}.

A useful description of gas capture at feeding scales is the hydrodynamic (HD) Bondi model \cite{bondi1952}, which assumes spherical, single-temperature gas accretion: 
\begin{equation}
    \mdotb =  4\pi \lambda_{\rm s} \dfrac{(GM_{\rm BH})^2}{c_{\rm s}^{3}} \rho_{0} = \pi \rb^2 \rho_{0} c_{\rm s} \ ,\
\label{eq:bondi_lambda}
\end{equation}
where $\rho_0,\ c_{\rm s}$ are the density and speed of sound far from the Bondi radius, $\rb=G\MBH/c_{\rm s}^2$,
and $\lambda_{\rm s}=1/4$ for nonrelativistic monatomic gas \cite{Shapiro1976}.
An ideal candidate for studying SMBH interactions with their hot, gas-rich environments is M87*, a nearby jetted AGN and key LLAGN \cite{forman2017}.  For M87*, $\rb \sim 0.1 \, \text{kpc}$, and the estimated inflow rate is, 
$\mdotb \sim 0.1 \, M_{\odot} \, \text{yr}^{-1}$ \cite{russell2015}.
The Event Horizon Telescope (EHT) has observed the innermost regions of the M87* accretion flow and constrained the BH mass accretion rate, $\mdotbh \simeq (3 - 20) \times 10^{-4} \, M_{\odot} \, \text{yr}^{-1}$~\cite{akiyama2021first}. This highlights the need for understanding the physical mechanisms, such as outflows, that allow the system to expel $ \gtrsim 99\% $ of the inflowing gas across $ \gtrsim 5 $ orders of magnitude in scale.

The Bondi model \cite{bondi1952} ignores both the magnetic fields and angular momentum and produces no outflows. HD models of RIAFs, such as the adiabatic inflow-outflow (ADIOS) model, find that outflows can result in mass-loss, which reduces the mass inflow rate as a power-law in radius, $\mdot\propto r^s$, with $ s \sim 0{-}1$ \cite{1999MNRAS.303L...1B, blandford_begelman2004, 2012MNRAS.420.2912B}. Simulations of rotating tori find $s \sim 0.4{-}0.75$ \cite{yuan2012numerical, yuan2012mhd}. 
Similar values are also found in simple convection accretion flow (SCAF) models, $s \simeq 0.7$ \cite{xu2023simple}. General relativistic magnetohydrodynamic (GRMHD) simulations of weakly magnetized accretion found weak outflows with $s \sim 1$ for non-spinning BHs~\cite{2012MNRAS.426.3241N,2015ApJ...804..101Y} and more powerful outflows with $s\sim0.5$ for spinning BHs~\cite{2013MNRAS.436.3856S}.
When large-scale poloidal fields accumulate, they can become dynamically-important, obstruct the accretion, and lead to a magnetically arrested disk (MAD) state \cite{1974Ap&SS..28...45B, 1976Ap&SS..42..401B, igu03, nia03, 2008ApJ...677..317I} that launches powerful jets with energy efficiencies exceeding $100$\% \cite{tchekhovskoy2011efficient, mckinney2012general}. Still, MAD simulations typically yield $s \approx 0.4{-}0.5$ \cite{2013MNRAS.436.3856S, 2024ApJ...965..175M}. 

A challenge in measuring $s$ in a rotating torus setup is the limited mass supply and the difficulty of attaining a steady state. In contrast, infinite mass reservoir approaches, e.g., realistic galaxy or Bondi-like configurations allow sustained BH fueling from an infinitely large ambient medium reservoir, and can attain long-term steady state solutions.
Recent studies have advanced the efforts to connect galactic and BH scales using methods such as super-Lagrangian refinement \cite{2021ApJ...917...53A, hopkins2023forge, hopkins2025zooming}, remapping between simulations~\cite{Kaaz_2025},  and nested mesh techniques \cite{ressler2018hydrodynamic, 2020ApJ...896L...6R, guo2023toward, guo2024mag}. Multi-zone approaches have achieved quasi-steady accretion flows onto non-spinning BHs out to $\rb \lesssim 10^7 \rg$ \cite{cho2023bridging, cho2024mag}, and have recently incorporated jet feedback and spin effects on scales $\rb \gtrsim 10^5 \rg$ \cite{guo2025cyclic}, though the resulting jets were weak and short-lived. 
However, these methods either neglect the back-reaction of feedback on the inflow or lack the long-term jet stability needed to reveal steady-state trends in the presence of large-scale jet feedback.

We adopt a different approach that avoids assumptions inherent to multi-scale models—such as fixed time hierarchies or constrained flow variability—by directly simulating Bondi accretion onto a spinning BH with ambient magnetic fields and gas rotation. This approach enables the self-consistent formation of accretion disks and jets.
3D GRMHD studies of non-rotating gas accretion onto spinning BHs \cite{ressler2021magnetically, Lalakos2024} showed that such flows are unstable, and give way to a rocking accretion disk (RAD) state, where angular momentum inflow, and jetted outflows, become randomly oriented \cite{Lalakos2024} (hereafter \citetalias{Lalakos2024}). 

In contrast, including angular momentum—parametrized by the circularization radius, \rcirc, at which the accretion flow forms a disk—can stabilize the MAD state \cite{Lalakos2022}.
While sustained large-scale vertical magnetic flux on the BH is key for maintaining the MAD state, the role of angular momentum remains less understood. Recent work \cite{galishnikova2025strongly} (see also \cite{kwan2023effects, chan2025impact}) suggests that insufficient angular momentum prevents stable disk and jet formation, pointing to a critical circularization radius below which the flow cannot remain in a stable MAD state.

We carried out the longest-duration contiguous magnetized Bondi-like accretion flow simulations in GRMHD to date ($\lesssim 4 \times 10^6 \rg/c$) over the length scales of up to $\rb/\rg = 10^4$ to explore the stability and properties of steady-state MADs. We parametrically explore the dependence of the accretion flow properties on both the Bondi radius and circularization radius.
Crucially, our simulations fully capture the time-dependent jet feedback loop, by directly connecting the BH and Bondi scales. Unless stated otherwise, 
we use the units $G = M_{\rm BH} = c = 1$. Thus, length is measured in
units of $\rg$, and time in units of $\tg = \rg / c$. 

\section{Methods} \label{sec:Numerical setup}


{\it Numerical setup.} We perform our simulations using the GRMHD code {\sc h-amr}, which features GPU acceleration, adaptive mesh refinement (AMR), and local adaptive timestepping \cite{liska2022h}. We set up the grid in spherical Kerr-Schild coordinates, ($r$, $\theta$, $\varphi$), with a uniform radial grid in $\log r$ covering $0.83\rh \le r \le 10^6 \rg$; the grid includes six cells inside the event horizon of radius, $\rh= \rg(1+\sqrt{1-a^2})$, to ensure that the inner radial boundary is causally disconnected from the BH exterior. Here, $a$ is the dimensionless BH spin parameter ($-1\le{}a\le1$). The polar and azimuthal grids are uniform, covering the ranges $0 \le \theta \le \pi$ and $0 \le \varphi \le 2\pi$, respectively. We apply outflow boundary conditions in $r$-, transmissive in $\theta$-, and periodic in $\varphi$-directions \cite{liska2022h}. We adopt the base grid resolution of $N_r \times N_{\theta} \times N_{\varphi} = 448 \times 96 \times 192$ in the $r$-, $\theta$-, and $\varphi$-directions, respectively, with a single block resolution of $N^{\rm B}_r \times N^{\rm B}_{\theta} \times N^{\rm B}_{\varphi} = 56 \times 48 \times 48$. For $r \ge 6.5 \rg$, we use one level of static mesh refinement (SMR), to double the effective resolution to $896 \times 192 \times 384$. This resolution is sufficient to resolve MAD accretion dynamics near the BH \cite{porth2019event}; see also \cite{galishnikova2025strongly} for a convergence study in a similar setup.


{\it Model.} We immerse a \BH of mass $M_{\rm BH}$ in a uniform ambient medium of mass density $\rho =\rho_{0}$. We ignore the radiation effects (e.g., cooling), as appropriate for low-luminosity \BH accretion; this allows free scaling of $\rho_{0}$, which we set to $\rho_{0}=1$ in all our simulations. Inside the Bondi radius ($r<\rb$), we carve out a cavity to avoid imposing predefined conditions that can affect the long-term evolution. We explore a broad range of spatial scale separation, by varying the normalized Bondi radius, $\rb/\rg = \left\{100,\, 300,\, 1000,\, 3000,\, 10^4\right\}$.  We choose the ambient gas angular momentum by setting the circularization radius, for which we explore several values, $\rcirc/\rg = \left\{ 0,\, 30,\, 120,\, 300 \right\}$. We adopt solid body rotation on spheres by choosing the angular momentum of the form, $\ell( \theta) = \ell_0 \sin^2 \theta$~\cite{palit2019time}, which peaks at the equator and vanishes at the poles, where $\ell_0 = \sqrt{GM_{\rm BH} \rcirc}$.

The \BH spin vector can generally be misaligned with the gas angular momentum, but we assume perfect alignment to simplify the physics and better identify underlying trends. We set a high dimensionless spin magnitude, $a = 0.94$, to maximize jet power and its impact on gas dynamics. We use an ideal gas equation of state, $p_{\rm g}=(\Gamma-1)u_{\rm g}$, where $p_{\rm g}$ and $u_{\rm g}$ are the gas pressure and internal energy, respectively. In this work, we focus on monatomic nonrelativistic gas with an adiabatic index, $\Gamma=5/3$. 

At $r\gg \rb$ we impose a large-scale vertical lab-frame magnetic field aligned with the \BH spin, and modify the radial field component so that $B^r$ smoothly vanishes at $r=\rb$. For this, we adopt a covariant magnetic vector potential, $A_{\varphi} \propto {\rm max}[(r^2-\rb^2),0] \sin^2\theta$. 
We normalize the magnetic field strength by setting the thermal-to-magnetic pressure ratio, plasma $\beta = p_{\rm g}/p_{\rm m} = 100$ at $r\gg \rb$, where $p_{\rm m} = b^2/8\pi \equiv b^\mu b_\mu / 8\pi$ is the magnetic pressure, and $b^\mu$ is the contravariant comoving magnetic field vector. This ensures $\beta \geq 100$ throughout the domain, so the magnetic field is initially subdominant. To break the axial and midplane symmetries and seed the magneto-rotational instability (MRI; \cite{bal91}), we introduce 2\% random thermal pressure perturbations into the initial conditions. 

At every radius, $r$, we compute the mass accretion rate as the angle-integrated radial rest-mass flux towards the BH:
\begin{equation}
    \dot{M}(r) = - \oiint  \rho u^r \sqrt{-g}\der\theta \der\varphi\ ,\
    \label{eqn:mdot_def}
\end{equation}
where $g= \left|g_{\mu \nu}\right|$ is the determinant of the four-metric tensor, $u^\mu$ is the coordinate-frame contravariant proper four-velocity vector, and the Greek indices run from 0 to 3.

The stress-energy tensor,
\begin{equation}
    T^\mu_\lambda = \left( \rho c^2+ u_{\rm g} + p_{\rm g} + \frac{b^2}{4\pi} \right ) u^\mu u_\lambda + \left( p_{\rm g} + \frac{b^2}{8\pi} \right) \delta^\mu_\lambda - \frac{b^\mu b_\lambda}{4\pi},
    \label{eqn:TOT_fl}
\end{equation}
allows us to compute the energy flux towards the BH similar to eq.~\eqref{eqn:mdot_def},
\begin{equation}
    \dot{E}(r) =  \oiint T^r_t \sqrt{-g}\der\theta \der\varphi\ .
    \label{eqn:power}
\end{equation}
We define the outflow energy efficiency,
\begin{equation}\label{eqn:eta}
\eta= \dfrac{\mdot c^2 - \edot }{\langle \mdot \rangle_\tau c^2}\ ,\
\end{equation}
where $\langle \mdot \rangle_\tau$ is the rolling average of the mass accretion rate over the time interval of $\tau = 3000\tg$. This time interval is sufficiently long to average over the strong \mdot oscillations in the MAD state.

We measure the total power, $\dot{E}$, efficiency, $\eta$, and BH mass accretion rate at $r = 5\rg$, i.e., $\mdotbh = \dot{M}(r=5\rg) $, to avoid potential contamination by the density floors near the event horizon (see \citepalias{Lalakos2024} for the description of the floors). Because $\dot{E}$ and  \mdot are conserved and independent of radius in a steady state, the time-averaged values remain unaffected, with only minor shifts in temporal dependencies by $\Delta t \lesssim 5 \tg$, which is much shorter than our simulation sampling interval of $100 \tg$. 

We quantify the strength of the magnetic flux relative to the accreting gas ram pressure, by defining the dimensionless absolute \BH magnetic flux as,
\begin{equation}\label{eqn:phibh}
\phibh = \dfrac{1}{2\sqrt{\langle \mdotbh\rangle_{\tau} \rg^2 c}} \oiint \left | B^r \right | \sqrt{-g}\der\theta \der\varphi\ ,\
\end{equation}
where the integral represents the total absolute magnetic flux on the BH event horizon, $r=\rh$, and the denominator provides the normalization based on the mass accretion rate. We carry out the integral in eq.~\eqref{eqn:phibh} over the entire BH horizon, and use the prefactor of $1/2$ to convert it to one hemisphere.

\begin{figure*}[ht!]
\centering
\includegraphics[width=1\textwidth]{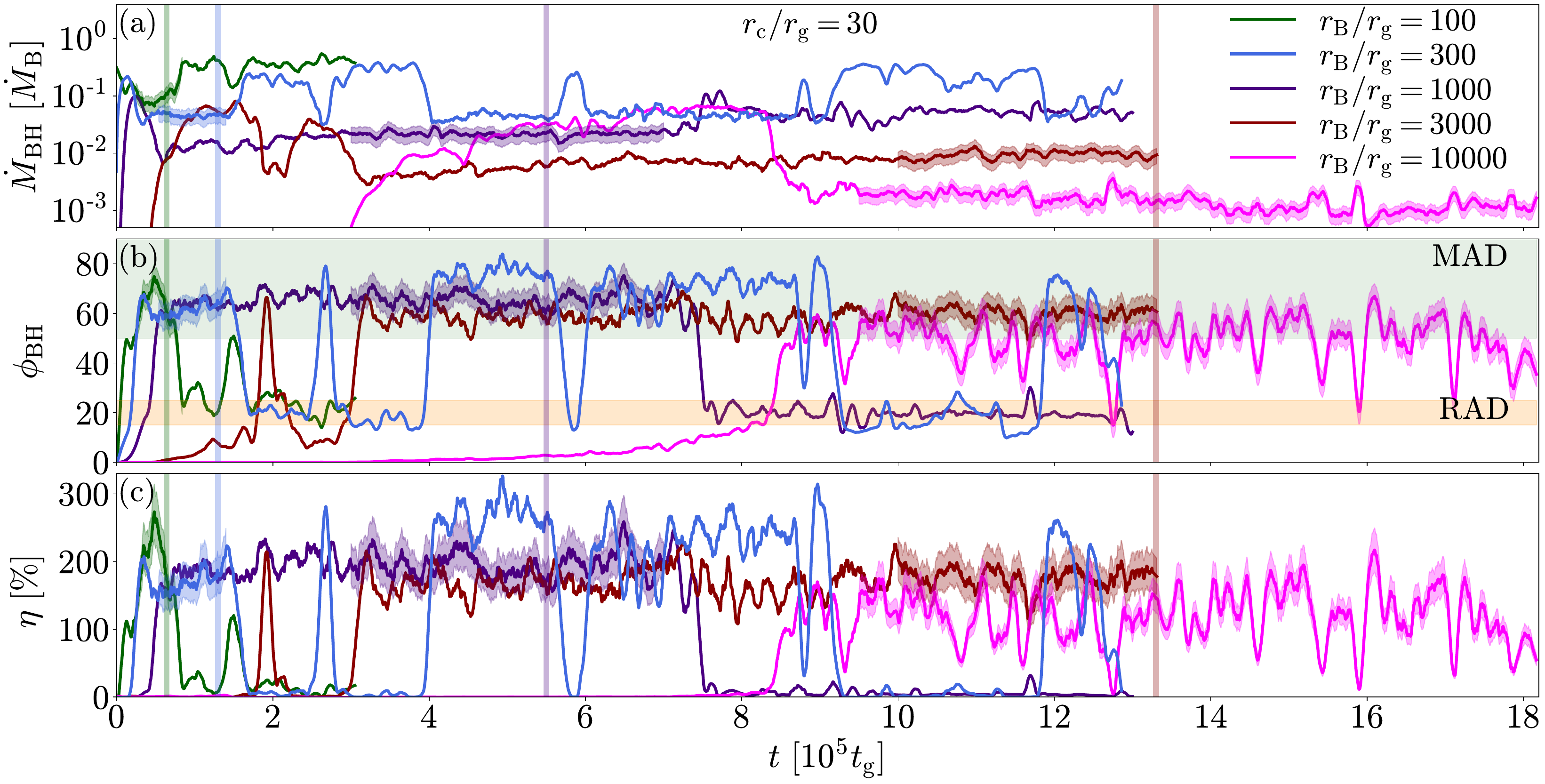}
\caption{
All of our simulations end up entering the MAD state, and some switch back and forth between the MAD (green shaded region) and RAD  (yellow shaded region)  states.
(a) Increasing the Bondi radius reduces BH mass accretion rates, \mdotbh, normalized to the Bondi accretion rate, \mdotb. 
(b) All runs enter the MAD state, with normalized magnetic flux $\phibh \gtrsim 50$ (green shaded region), and some reach the RAD state (orange shaded region), in which $\phibh \sim 20$, with jets and disk continuously reorienting themselves. 
(c) The outflow power exceeds the accretion power, $\eta \sim 200\%$, consistent with the MAD regime.  In the RAD state, outflow efficiency is lower, $1\% \lesssim \eta \lesssim 10\%$. The $\rb=10^4 \rg$ case is still in the initial transient state, with a tilted accretion disk struggling to provide magnetic flux to the BH \cite{chatterjee2023}. For $\rb=300\rg$, we observe multiple transitions between the MAD and RAD states, which sometimes reach extreme values, $\eta \sim 250\%$ and $\phibh \simeq 75$. For clarity, we smoothed all quantities over a timescale of $10^4\,\rg/c$ using a zeroth-order Savitzky–Golay filter.
}
\label{fig:mdot_vs_time}
\end{figure*}

\section{Results}
\subsection{Effects of the Bondi Radius} \label{sec:rbondi}

Here, we consider a range of Bondi radius values, $\rb/\rg = \left\{100,\, 300,\, 1000,\, 3000,\, 10^4\right\}$, for a fixed circularization radius, $\rcirc/\rg = 30$. Fig.~\ref{fig:mdot_vs_time}(a) shows that after each simulation starts, the gas reaches the BH roughly on the Bondi timescale, $\tb = \rb / c_{\rm s} = (\rb/\rg)^{3/2} \tg$, and the BH accretion rate peaks at $\mdotbh \lesssim \mdotb$ soon thereafter.
Fig.~\ref{fig:mdot_vs_time}(b) shows that once the dimensionless BH magnetic flux, $\phibh$ (eq.~\ref{eqn:phibh}), becomes large enough, $\phibh \gtrsim 15$, the BH launches jets, with an outflow energy efficiency, $\eta \sim 10\%$ (Fig.~\ref{fig:mdot_vs_time}c). The jets clear out the polar funnel region and drive a blast wave into the gas.
Following the jet onset, $\mdotbh/\mdotb$ reaches its minimum value and then gradually increases. 
Fig.~\ref{fig:mdot_vs_time}(b,c) shows that once $\phibh \gtrsim 50$, the outflow efficiency, $\eta\sim200\%$, well exceeds $100$\%, both of which are consistent with the MAD state~\cite{tchekhovskoy2011efficient}. We find that all runs eventually enter the MAD state ($\phibh \gtrsim 50$, rectangular light-green shaded region), given sufficiently long simulation duration ($t \sim \text{few} \times \tb$) and sufficiently large ambient magnetic flux reservoir. 

\begin{figure*}[ht!]
\centering
\includegraphics[width=1\textwidth]{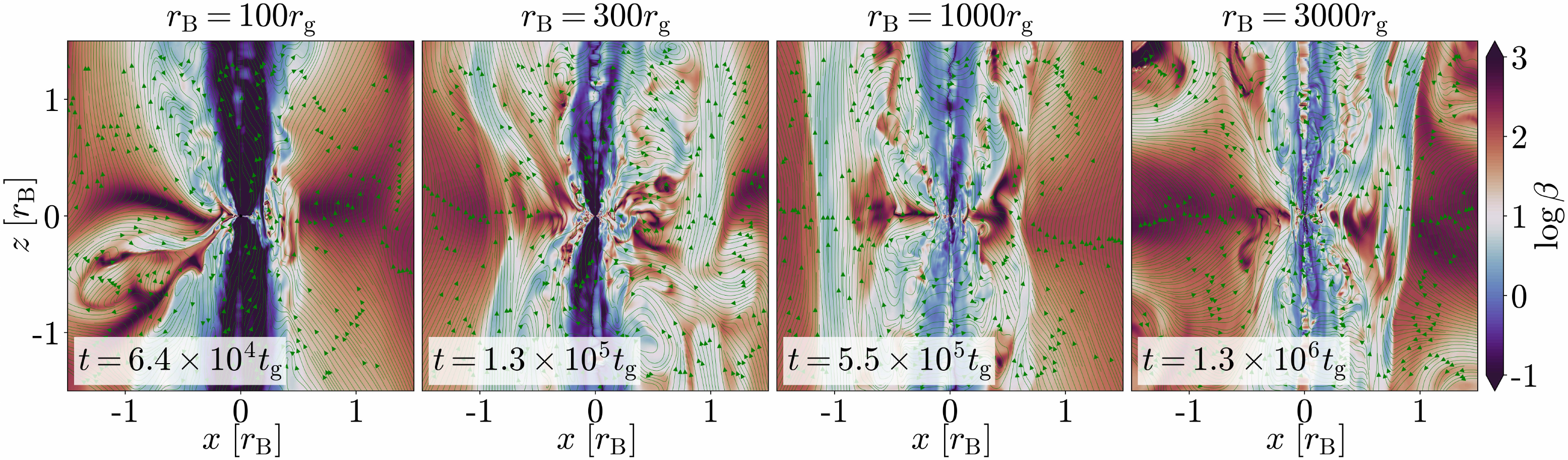}
\caption{The MAD state features erupting bundles of magnetic flux near the BH, visible as vertically directed field lines (green), which tear through the disk and can extend out to the Bondi radius.
The panels show meridional slices ($3 \times \rb$ on the side) through the logarithm of the plasma-$\beta$ parameter (blue: magnetic-pressure dominated, red: thermal-pressure dominated), at the indicated times, $t/\tg$, and scale separations, $\rb/\rg$. To facilitate comparison, we show the outcome of the simulations with the same $\rcirc/\rg=30$ in all panels. 
}
\label{fig:panels}
\end{figure*}

Fig.~\ref{fig:panels} presents snapshots of each of our $\rcirc/\rg = 30$ simulations in the MAD state. 
The color shows the logarithm of plasma $\beta$, where blue indicates magnetic-pressure dominated regions ($\beta < 1$) and red represents thermal-pressure dominated regions ($\beta > 1$). Green directed lines show the poloidal magnetic field lines traced out in the $x{-}z$ plane. 
The magnetic flux eruptions, which have $\log \beta\sim 0.5$, occasionally escape from the BH and rip through the disk, as characteristic of MAD flows \cite{Ripperda_2020, Ripperda_2022}. We find that the eruptions can buoyantly rise and reach $r=\rb$.

When run sufficiently long, many of our models exit the MAD state, as indicated by a sudden drop in $\phibh$ to $\sim20$: this signals a transition to the RAD state (orange shaded region). For example, in the $\rb/\rg = 300$ run, at $t\simeq 1.5 \times 10^5 \tg$, the dimensionless magnetic flux $\phibh$ drops from $\sim60$ to $\sim20$, and $\mdotbh$ increases by nearly an order of magnitude, from $\avmdotbh/\mdotb \simeq 5\times 10^{-2}$ to $ \simeq 0.3$. 
For $\rb/\rg = 1000$, the MAD state lasts from $0.5 \times 10^5 \lesssim t/\tg \lesssim 7.5 \times 10^5$, after which the flow turns RAD. For $\rb/\rg = 3000$ and $10^4$, we do not observe the transition to the RAD state, most likely due to the limited duration of these simulations.

In the RAD state, the outflow efficiency $\eta$ ranges from $1\%$ to $10\%$, with higher values corresponding to brief jet episodes, and lower values occurring when jets fail to form due to flux cancellation or disk tilt. The average BH magnetic flux remains in the range $15 \lesssim \phibh \lesssim 25$: near the lower end, jets shut off and $\mdotbh$ peaks; near the upper end, weak, fluctuating jets emerge, with $\eta \lesssim 10\%$ and moderately reduced $\mdotbh$, still exceeding MAD levels.

Here, we point out that for our largest-scale separation simulation, with $\rb/\rg = 10^4$, $\phibh$ fluctuates between $\sim 20$ and $\sim 60$. The simulation has not yet reached a stable midplane-aligned disk and instead forms a tilted disk \cite{liska2018formation}. This misalignment disrupts the coherent advection of magnetic flux needed to sustain a steady MAD state \cite{chatterjee2023} and causes the jets to bend as they follow the rotational axis of the tilted accretion flow (see Fig.~\ref{fig:tilted_disk} in Appendix~\ref{appendixA}). 
The duration from jet onset to the end of the run spans roughly one Bondi time ($t \sim \tb$), suggesting that longer evolution times ($t \gtrsim \text{few} \times \tb$) may lead to a stable, aligned MAD state.

\citetalias{Lalakos2024} studied the MAD-to-RAD transition in a system where gas at Bondi scales had zero angular momentum, with the MAD state lasting $t\simeq 5\times 10^4 \tg$. Here, even with ambient angular momentum, $\rcirc=30\rg$, we observe the RAD transition, however, occurs at one order of magnitude longer timescales.
Interestingly, a cyclic MAD-to-RAD transition is evident for $\rb/\rg = 300$: for example, the system re-enters the MAD state thrice, at $t/\tg \simeq \{4, 9, 12\} \times 10^5$. Some RAD-to-MAD transitions are followed by a boost in magnetic flux, $\phibh \sim 75$, e.g., at $4\times 10^5 \lesssim t/\tg \lesssim 5.5\times 10^5$. The resulting MAD state produces significantly more powerful outflows, with average energy efficiencies reaching $\eta \sim 250\%$. This suggests that RAD–to-MAD transitions may not only restore jet activity but also amplify it beyond prior levels.

Our $\rb/\rg = 300$ simulation is closely related to the $\rb/\rg = 250$ case in \cite{galishnikova2025strongly}, who define $\rb$ twice as large, labeling it as $\rb/\rg = 500$. They identify this run as their most stable run and suggest that larger $\rb$ values lead to increased variability. Indeed, their $\rb/\rg = 250$ simulation spans multiple Bondi timescales and maintains a MAD state for durations comparable to those observed in our MAD–RAD duty cycle. However, for simulations with larger scale separations, their evolution time might be too short to assess the emergence of a MAD state. We find that our simulations robustly form the MAD state, given sufficient (i) simulation duration (several Bondi times) and (ii) supply of the large-scale vertical magnetic flux at the Bondi scales.

\subsection{ MAD-RAD Inflow Scaling} 

Understanding what fraction of \mdotb ultimately reaches the BH, particularly in the presence of jet and outflow feedback, and across multiple orders of magnitude in radius, requires long-duration simulations. 
Additionally, the angular momentum content of the gas, parameterized by the circularization radius, \rcirc, can also influence this fraction: e.g., larger \rcirc values may result in stronger and more extended outflows, and, potentially, in a delayed onset of the RAD state.

\begin{figure}[!t]
\centering
\includegraphics[width=1.0\columnwidth]{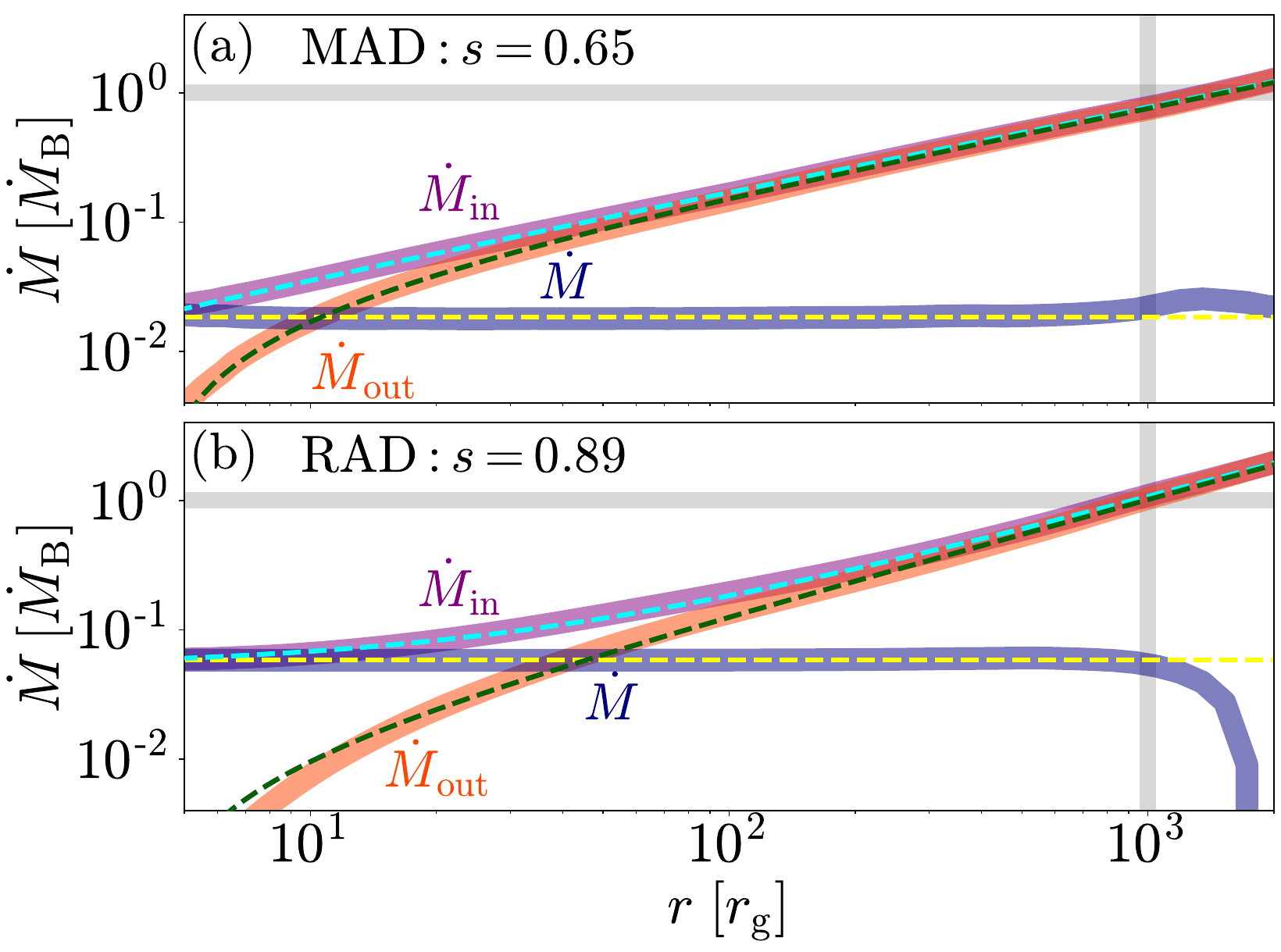}
\caption{
We can accurately determine the radial profiles of MAD and RAD states because both of them attain the inflow equilibrium out to the Bondi radius for our $\rb=1000\rg$ simulations ($\rcirc=120\rg$ for the MAD and $\rcirc=30\rg$ for the RAD plot). The mass inflow accretion scaling in the MAD state is $\mdotin \sim r^{0.65}$, while the RAD state has a steeper slope, $\mdotin \sim r^{0.89}$. The RAD outflow rate, $\mdotout$, is lower near the BH, which results in the inflow rate to transition from a power-law to a constant value at larger radii, $r\simeq 50\rg$, compared to the MAD state, which happens at $r\simeq 10\rg$.
}
\label{fig:inflow_eq}
\end{figure}

Fig.~\ref{fig:inflow_eq} shows the time-average inflow ($\mdotin$, purple), outflow ($\mdotout$, orange), and net ($\mdot$, blue) mass accretion rates for both MAD and RAD states. Fig.~\ref{fig:inflow_eq}(a) shows the MAD state averaged over the longest MAD episode in the $\rb/\rg=1000$, $\rcirc/\rg=120$ simulation whose inflow equilibrium region, in which the net mass accretion rate is conserved, $\mdotin-\mdotout\equiv \mdot = \text{constant}$, extends out to $r=\rb$. At large radii, $r \gg 10\rg$, the inflow and outflow rates are similar and follow the same power-law dependence, $\mdotin \approx \mdotout \propto r^s$ with $s=0.65$. At small radii, $r \lesssim 10\rg$, the inflow rate flattens out, $\mdotin \simeq \mdot$, and the outflow rate vanishes, $\mdotout \ll \mdot$.
This motivates us to describe the outflow and inflow rates as follows:  
\begin{align}
    \mdotout^\text{fit}(r) &= \mdotbh  
    \times \max\left(\dfrac{r^s - \rw^s}{r_0^s - \rw^s},0\right)
    ,\
    \label{eqn:modelout}\\
    \mdotin^\text{fit}(r) &= \mdotbh + \mdotout^\text{fit}(r)
    ,\
    \label{eqn:modelin}
\end{align}
where $\rw$ is the wind-launching radius, such that $\mdotout^\text{fit}(\rw)=0$, and $r_0 \simeq 10\rg$ is the distance at which the outflow becomes important, such that $\mdotout^\text{fit}(r_0)=\mdotbh$. When using eq.~\eqref{eqn:modelout} to fit the simulation results, we consider only $r \ge 5\rg$, to exclude the density floor region near the BH. 
Note that eqs.~\eqref{eqn:modelout} and~\eqref{eqn:modelin} do a good job at capturing the curvature of $\mdotout$ and $\mdotin$ curves in Fig.~\ref{fig:inflow_eq}, although small deviations may persist near $r = 5\rg$. As a result, the best-fit value of the power-law index, $s$, is sensitive to neither the distance, $r$, at which we fit the simulation data nor to whether we fit the inflow or outflow rate. 

In contrast, using the local logarithmic slope $s_{\#} \equiv \der\log\mdot_{\#}/\der\log r$ biases estimates: it underestimates the inflow slope ($s_\text{in} \sim 0.5$) due to $\mdotin(r)$ flattening near $r_0$, and overestimates the outflow slope ($s_\text{out} \sim 1$) due to $\mdotout(r)$ steepening there. This may partly explain the wide range of slopes reported in simulations, $0.5\lesssim s\lesssim1$ \citep{mckinney2012general,2012MNRAS.426.3241N,2013MNRAS.436.3856S,2015ApJ...804..101Y}.

In the RAD state, Fig.~\ref{fig:inflow_eq}(b) shows the results of the $\rb/\rg=1000$ and $\rcirc/\rg = 30$ simulation, which has the longest RAD steady-state, which reaches inflow equilibrium out to \rb. We find a steeper slope, $s =0.89$. The outflow rate is on average weaker than the MAD state, as $\mdotout = \mdotbh$ at $r_0 \simeq 50 \rg$, which is where the inflow rate starts flattening.

For simplicity, we have so far shown time-averaged radial profiles for only two simulations. In Appendix~\ref{appendixB}, we extend this analysis to additional runs: we demonstrate that averaged inflow profiles are independent of \rb (Fig.~\ref{fig:all_rb}), and changes in \rcirc affect neither the time-averaged inflow profiles (Fig.~\ref{fig:all_rc}) nor the temporal evolution near the BH (Fig.~\ref{fig:mdot_vs_rcirc}). Therefore, we find no need to include \rcirc as a parameter into our description of the simulated radial profiles. 
However, larger $\rcirc$ values appear to delay the onset of the RAD state. For instance (see Fig.~\ref{fig:mdot_vs_rcirc}), the $\rcirc/\rg = 30$ simulation transitions to the RAD state after $t/\tg \simeq 7\times 10^5$, while the $\rcirc/\rg = 300$ case remains in the MAD state until $t/\tg \simeq 1.3\times 10^6$. Some of this variation may also reflect stochastic behavior in the turbulent flow.

\begin{figure}[!t]
\centering
\includegraphics[width=1\columnwidth]{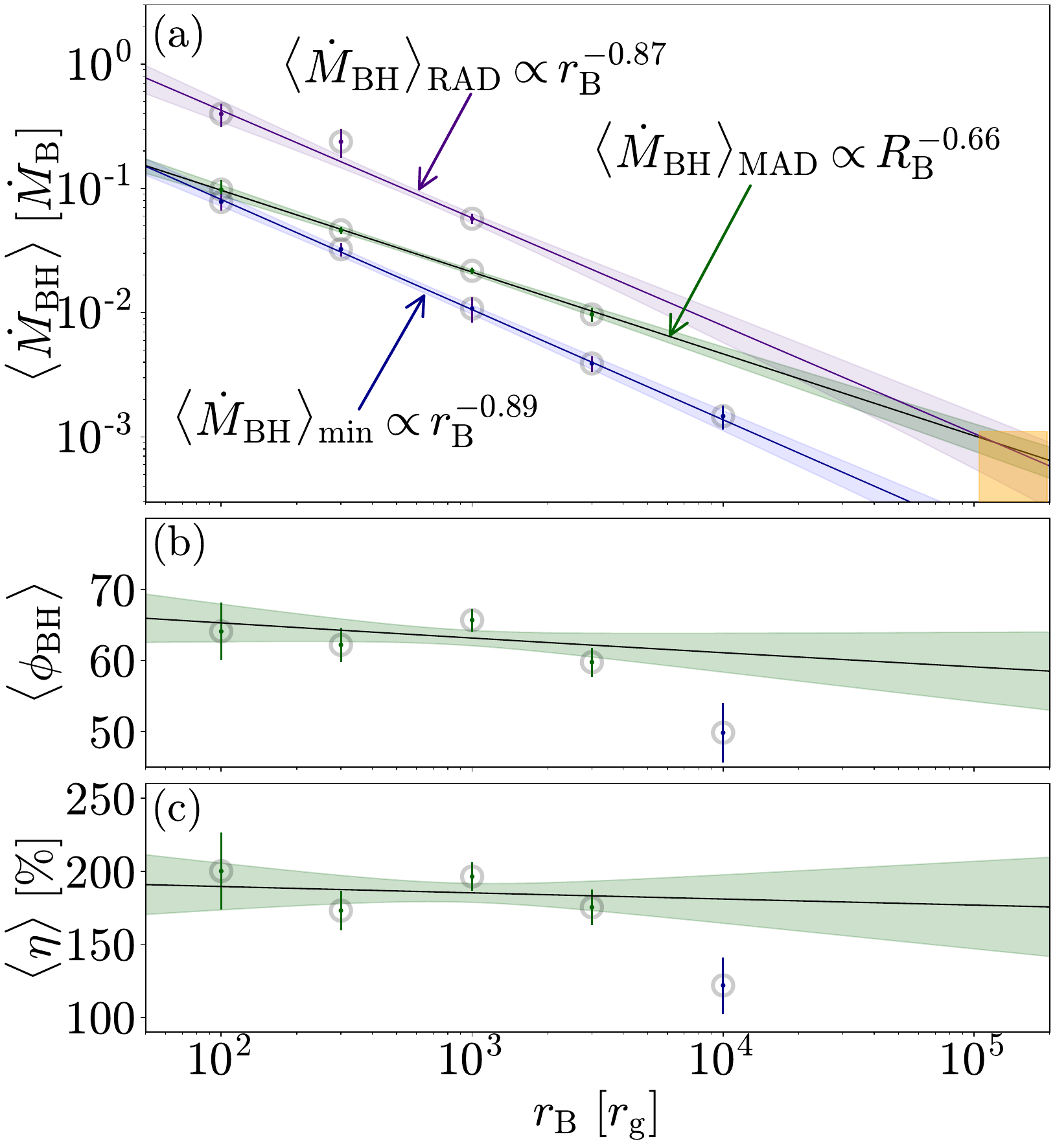}
\caption{
(a) Time averaged normalized mass accretion rate in the MAD state, $\avmdotbh/\mdotb$ (green), decreases with increasing $\rb/\rg$, following a power-law with the best-fit index of $(-0.66)$, for the set of our simulations with $\rcirc/\rg=30$. Vertical error bars indicate $2\sigma$ uncertainties in the data points, while the shaded region shows the $2\sigma$ confidence interval of the fit. The data point for $\rb/\rg = 10^4$ is excluded from the fit, as it corresponds to the transient, early jet onset phase (blue), characterized by a steeper index of $(-0.89)$. We anticipate that the $\rb/\rg = 10^4$ simulation will eventually transition to the MAD state (blue to green). For the RAD state (purple), the best-fit power-law index is $-0.87$.
(b, c) Both the average BH magnetic flux, $\langle \phi_{\rm BH} \rangle \approx 65$, and outflow efficiency, $\langle \eta \rangle \approx 200\%$, remain approximately constant across $\rb$. 
The MAD and RAD fits converge at $\rb/\rg \gtrsim 10^5$ (orange shaded region).
}
\label{fig:mdot_vs_rb}
\end{figure}


\subsection{Universal scaling relations} \label{sec:mdot_powerlaw}
Here, we examine how the time-average BH mass accretion rate in the MAD state depends on the Bondi radius.
In Fig.~\ref{fig:all_rb} in Appendix~\ref{appendixB}, we show time-averaged inflow radial profiles, demonstrating that the MAD inflow equilibrium extends out to $\rb$, for $\rb \le 1000\rg$, and out to $\sim 0.5\rb$ for $\rb=3000\rg$. 
The time-averaging windows used for each $\rb/\rg$ simulation are indicated by the shaded regions in Fig.~\ref{fig:mdot_vs_time}.
To account for potential data correlations in time, we calculate the error of the averaged values as detailed in Appendix~\ref{appendixc}.
We choose a time-averaging interval of sufficient duration, \( \Delta t \gtrsim \tb \), that starts after the jets launch and $\mdotbh$ settles onto a steady state value.

Fig.~\ref{fig:mdot_vs_rb}(a) presents the time-average values of $\langle \mdotbh \rangle /\mdotb$ in the MAD state, as a function of $\rb/\rg$ (green), with $2\sigma$ error bars and shaded confidence interval. We fit the data using a power-law fit, excluding $\rb/\rg = 10^4$ which has not reached a steady state:  
\begin{equation} \label{eq:mdot_scaling}
 \text{MAD:}\quad\dfrac{\avmdotbh}{\mdotb} = (1.0 \pm 0.2) \times 10^{-3} \times \left( \dfrac{\rb}{10^5\rg}\right)^{-0.66\pm 0.03}.
\end{equation}

In Fig.~\ref{fig:mdot_vs_rb}(b), we show the time-averaged normalized magnetic flux, $ \avphibh$, as a function of the Bondi radius, $\rb$. The best-fit line has a slope consistent with zero, $ \avphibh = (70 \pm 6) \times (\rb /\rg)^{-0.01 \pm 0.01}$, which implies that it most likely is independent of the scale separation. Fig.~\ref{fig:mdot_vs_rb}(c) shows that the average outflow efficiency also has a slope consistent with zero, $\aveta = (200 \%\pm 20\%) \times (\rb /\rg)^{-0.01 \pm 0.01}$, which is consistent with the energy outflow efficiency exceeding $100$\% in MADs~\citep{tchekhovskoy2011efficient}.

Fig.~\ref{fig:mdot_vs_rb}(a) additionally shows the accretion rate during the early jet launching phase of each simulation, $\avmdotbh_{\raisebox{-2pt}{\footnotesize min}}/\mdotb$ (blue), which features a steeper best-fit power-law index, $s = -0.89 \pm 0.03$. Interestingly, our $\rb/\rg=10^4$ simulation data point falls right on this scaling: this indicates that at later times ($t\gtrsim \tb$) this run can eventually fall on the extrapolation of our MAD-state state fit (green).

Similarly, for the RAD state (purple), we find: 
\begin{equation} \label{eq:rad_scaling}
    \text{RAD:} \quad \dfrac{\avmdotbh}{\mdotb} = (1.0 \pm 0.3) \times 10^{-3} \times \left( \dfrac{\rb}{10^5\rg}\right)^{-0.87\pm 0.05}\ .
\end{equation}
The RAD state represents a frustrated flow that is sufficiently magnetized to partially impede accretion ($15 < \avphibh < 25$), but without the strong, stable jets characteristic of MAD.

If the above power-law fits for $\avmdotbh$ vs $\rb$ for MAD and RAD states, eqs.~\eqref{eq:mdot_scaling} and~\eqref{eq:rad_scaling}, respectively, persist to larger scale separations, the two fits intersect at $\rb/\rg \sim 10^5$ (orange shaded region). This suggests that for realistic scale separations, $\rb/\rg\sim10^5{-}10^6$, the system might transition between MAD and RAD states without significant changes in \mdotbh.
Also, since RAD jets are intermittent, it remains unclear under what conditions jets at $\rb/\rg \gtrsim 10^5$ stay stable and aligned (e.g., M87*) or become disrupted (e.g., Sgr A*).


\section{Conclusions}\label{sec:conclusions}
We have presented a comprehensive study of hot accretion flows onto rapidly spinning BHs ($a=0.94$), using global 3D GRMHD simulations that span extreme scale separations and long contiguous durations.
We vary the Bondi-to-horizon scale separation, $\rb/\rg$, and gas rotation rate, parameterized by the dimensionless circularization radius, $\rcirc/\rg$, to assess their impact on the coupling between feeding and near-BH scales in LLAGN.
All runs use $\beta = 100$ at $r \ge \rb$; variations in $\beta$ have minimal impact on the steady-state near-BH accretion flow \cite{galishnikova2025strongly}.

All our simulations reach the MAD state once sufficient magnetic flux accumulates on the BH, with an average normalized BH magnetic flux $\langle \phibh \rangle \simeq 65$, and outflow energy efficiency, $\langle \eta \rangle \simeq 200\%$, i.e.\hbox{} the outflow power is twice the accretion power, $\mdotbh c^2$. In the MAD state, the outflow feedback and magnetic flux eruptions set the BH accretion rate and the steady-state inflow equilibrium region expands inside out. We find a universal power-law scaling $\avmdotbh/\mdotb \sim (\rb/\rg)^{-s}$ with $s=0.66\pm 0.03$, which is independent of \rb and \rcirc.
This slope agrees with the range of slopes found in hydrodynamic ADIOS~\cite{1999MNRAS.303L...1B, blandford_begelman2004, 2012MNRAS.420.2912B} and SCAF~\cite{xu2023simple} solutions, and is steeper than $s= 0.5$, reported in many (GR)MHD simulations  \cite[with some including jet feedback][]{ressler2020, ressler2021magnetically, cho2023bridging, cho2024mag, guo2025cyclic}. We note that including radiation effects and gas cooling can significantly alter the dynamics of the accreting flows \cite{rohoza2024turn, guo2024mag} and affect the radial scalings. Our findings are potentially relevant for accretion onto binary BHs, which show similar signs of jet and outflow feedback on the accretion in the MAD state \cite{Most:2024qus,ressler2025dual, wang2025galactic}.

Jet feedback at the feeding scale ($\rb$) scrambles angular momentum and naturally triggers the rocking accretion disk (RAD) state \cite{Lalakos2024}. Despite significant ambient rotation, RADs emerge robustly and often persist as long as—or longer than—MADs. RADs show reduced BH magnetic flux, $\langle \phibh \rangle \sim 20$, and efficiency, $\langle \eta \rangle \lesssim 10\%$, with jets that reorient and disrupt within the Bondi radius. Mass inflow accretion slope in RADs is steeper than in MADs, with $s = 0.87 \pm 0.05$, also independent of $\rb$ and $\rcirc$. Outflows are weaker, with the inflow accretion rate flattening at larger radii, $r_0 \simeq 50\rg$, compared to $r_0 \simeq 10\rg$ in MADs.

A larger circularization radius can delay the onset of RADs by preserving coherent angular momentum, potentially explaining the stability of long-lived, collimated jets \cite{oei2024black}. At typical scale separations ($\rb/\rg \gtrsim 10^5$), MAD and RAD accretion rates converge, raising the possibility that transitions between these states could occur without significant changes in $\mdotbh$, and that jets may either remain stable (MAD-like) or wobble intermittently (RAD-like).

Some simulations exhibit a MAD–RAD duty cycle that lasts from a few to tens of Bondi timescales,
$\tb \sim 0.2\,\text{Myr} \times (\rb/10^5\rg)^{3/2} \times (\MBH/10^9M_\odot)$.
For $\rb/\rg = 300$, $\tb/\tg \sim 5200$ is much shorter than the observed durations of the MAD and RAD states, $\tau \sim 10^5 \tg$. 
For M87*, this implies $\tau \sim 4\,\text{Myr}$, consistent with observations \cite{forman2017}. These findings suggest that duty cycles in LLAGN, governed by the MAD–RAD transition, may naturally explain the observed variability and longevity of AGN outbursts.


\begin{acknowledgments}
We thank Martijn Oei, Ioannis Liodakis, Anthony Readhead, and Eliot Quataert for helpful discussions.
AT was supported by 
NSF grants
AST-2009884,
AST-2107839,
AST-1815304,
AST-1911080,
AST-2206471, 
AST-2407475,
and
OAC-2031997, and by NASA grants 
80NSSC22K0031,
80NSSC22K0799,
80NSSC18K0565,
and 80NSSC21K1746.
Support for this work was provided by the National Aeronautics and Space Administration through \chandra Award Number TM1-22005X issued by the \chandra X-ray Center, which is operated by the Smithsonian Astrophysical Observatory for and on behalf of the National Aeronautics Space Administration under contract NAS8-03060.
The authors acknowledge the Texas Advanced Computing Center (TACC) at
The University of Texas at Austin for providing HPC and visualization
resources that have contributed to the research results reported
within this paper via the LRAC allocation AST20011
(\url{http://www.tacc.utexas.edu}). 

This research used resources from the Oak Ridge Leadership Computing Facility, which is a DOE Office of Science User Facility supported under Contract DE-AC05-00OR22725. An award of computer time was provided by the ASCR Leadership Computing Challenge (ALCC), Innovative and Novel Computational Impact on Theory and Experiment (INCITE), and the OLCF Director's Discretionary Allocation programs under award PHY129. This research used resources of the National Energy Research Scientific Computing Center, a DOE Office of Science User Facility supported by the Office of Science of the U.S. Department of Energy under Contract No. DE-AC02-05CH11231 using NERSC award ALCC-ERCAP0022634. This research used resources of the National Energy Research Scientific Computing Center (NERSC), a U.S. Department of Energy Office of Science User Facility located at Lawrence Berkeley National Laboratory, operated under Contract No. DE-AC02-05CH11231 using NERSC ERCAP award m2401 for 2022 and 2023.

ERM acknowledges support through the Oak Ridge Leadership Computing Facility at the Oak Ridge National Laboratory, which is supported by the Office of Science of the U.S. Department of Energy under Contract No. DE-AC05-00OR22725, and the National Energy Research
Scientific Computing Center, a DOE Office of Science User Facility
supported by the Office of Science of the U.S. Department of Energy
under Contract No. DE-AC02-05CH11231 and NERSC award
NP-ERCAP0028480.
BR acknowledges support by the Natural Sciences \& Engineering Research Council of Canada (NSERC), the Canadian Space Agency (23JWGO2A01), and by a grant from the Simons Foundation (MP-SCMPS-00001470). BR acknowledges a guest researcher position at the Flatiron Institute, supported by the Simons Foundation. 

\end{acknowledgments}


\appendix

\section{Misaligned flow forms precessing Disk}
\label{appendixA}

\begin{figure}[h]
\centering
\includegraphics[width=\columnwidth]{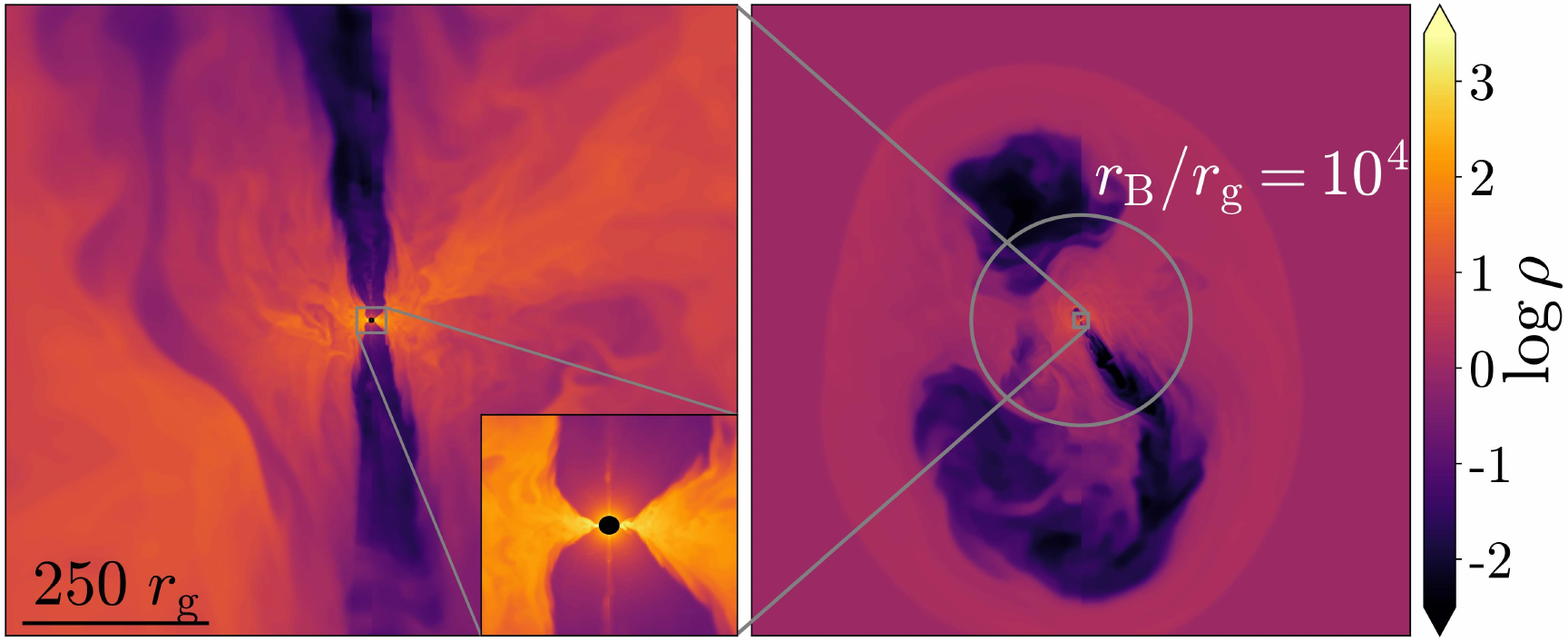}
\caption{
In our largest scale separation run ($\rb/\rg=10^4$, $\rcirc/\rg=30$), the accretion flow remains misaligned by $\sim45^\circ$ relative to the BH axis, shown at $t/\tg \sim 1.8\times10^6$. The system has not reached a MAD steady state, and the jets, significantly bent by the disk tilt, barely extend beyond the Bondi radius. Color shows contours of the logarithm of density, $\log \rho$. The left and right panels span $1000\rg$ and $6\times10^4\rg$, respectively, with the $40\rg$ inset showing the inner MAD aligned with the equatorial plane \cite{liska2018formation}. 
}
\label{fig:tilted_disk}
\end{figure}

\section{Effects of Circularization Radius} 
\label{appendixB}

\begin{figure}[h]
\centering
\includegraphics[width=\columnwidth]{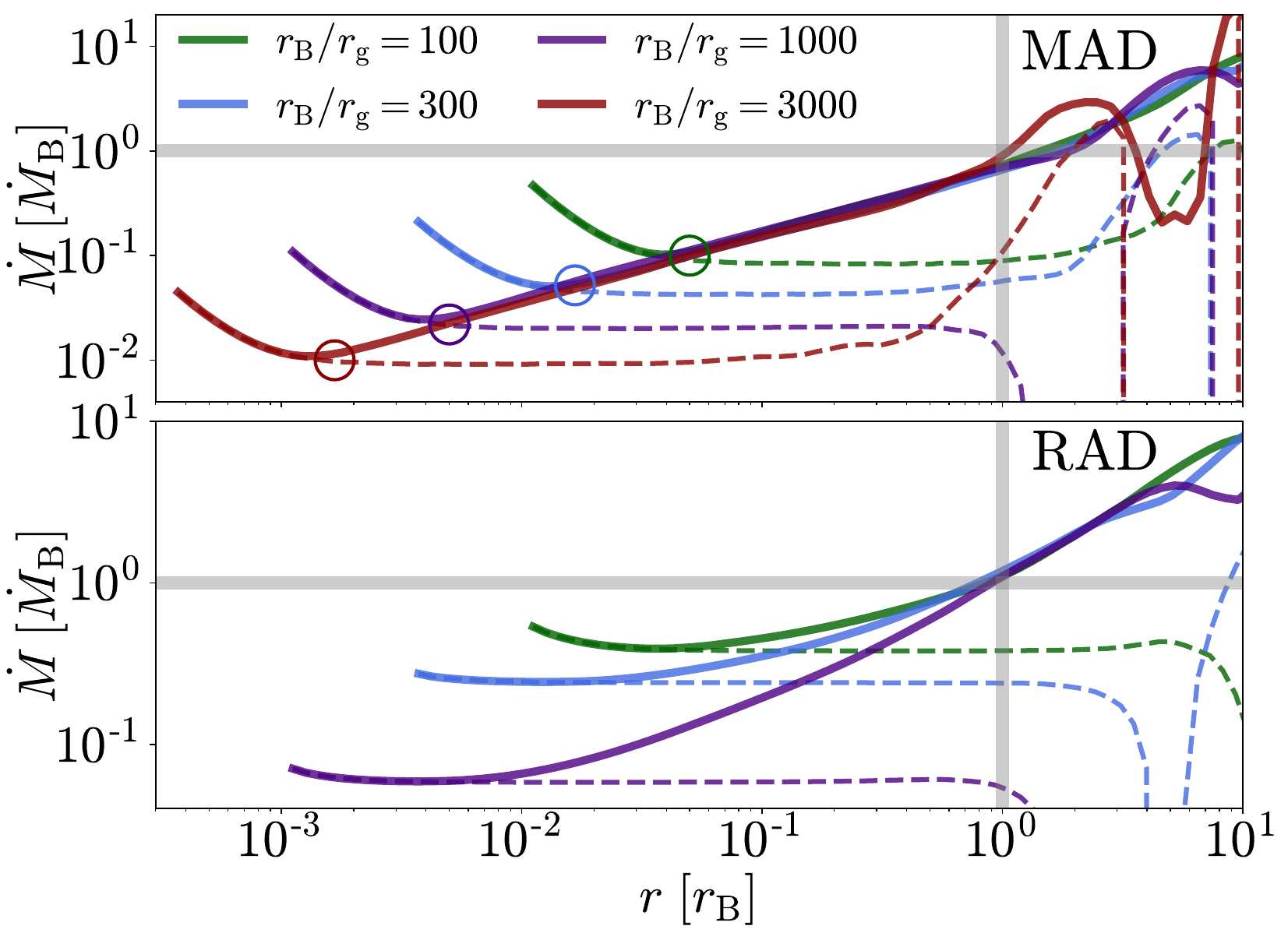}
\caption{
We find that magnetized rotating Bondi flows universally satisfy $\mdotin(\rb) \approx \mdotb$ and follow a single power-law profile, $\mdotin \propto r^s$, down to the wind base at $r = r_0$, where the profile flattens to $\mdotin \rightarrow \mdotb$. The parameters $r_0/\rg \ll \rb/\rg$ and $s$ are intrinsic to the flow and independent of the scale separation $\rb/\rg$, which only determines the radial extent of the power-law region and the minimum value of $\mdotin$.
The simple power-law scaling of the MAD ($s \sim 0.65$) and RAD ($s \sim 0.9$) states remains pinned down at $r=\rb$, $\mdotin\approx \mdotb$, and extends down as a single power law to the BH, where it flattens out at the same distances, $r_0$ (in units of $\rg$). 
Colored circles indicate where $\mdotbh$ (dashed lines) is measured at $r = 5 \rg$ to avoid contamination from density floors. The turnover from power-law (\mdotin, solid lines) to \mdot-constant profile occurs at larger radii for smaller $\rb/\rg$, as if the BH were effectively larger in size. 
}
\label{fig:all_rb}
\end{figure}

\begin{figure}
\centering
\includegraphics[width=\columnwidth]{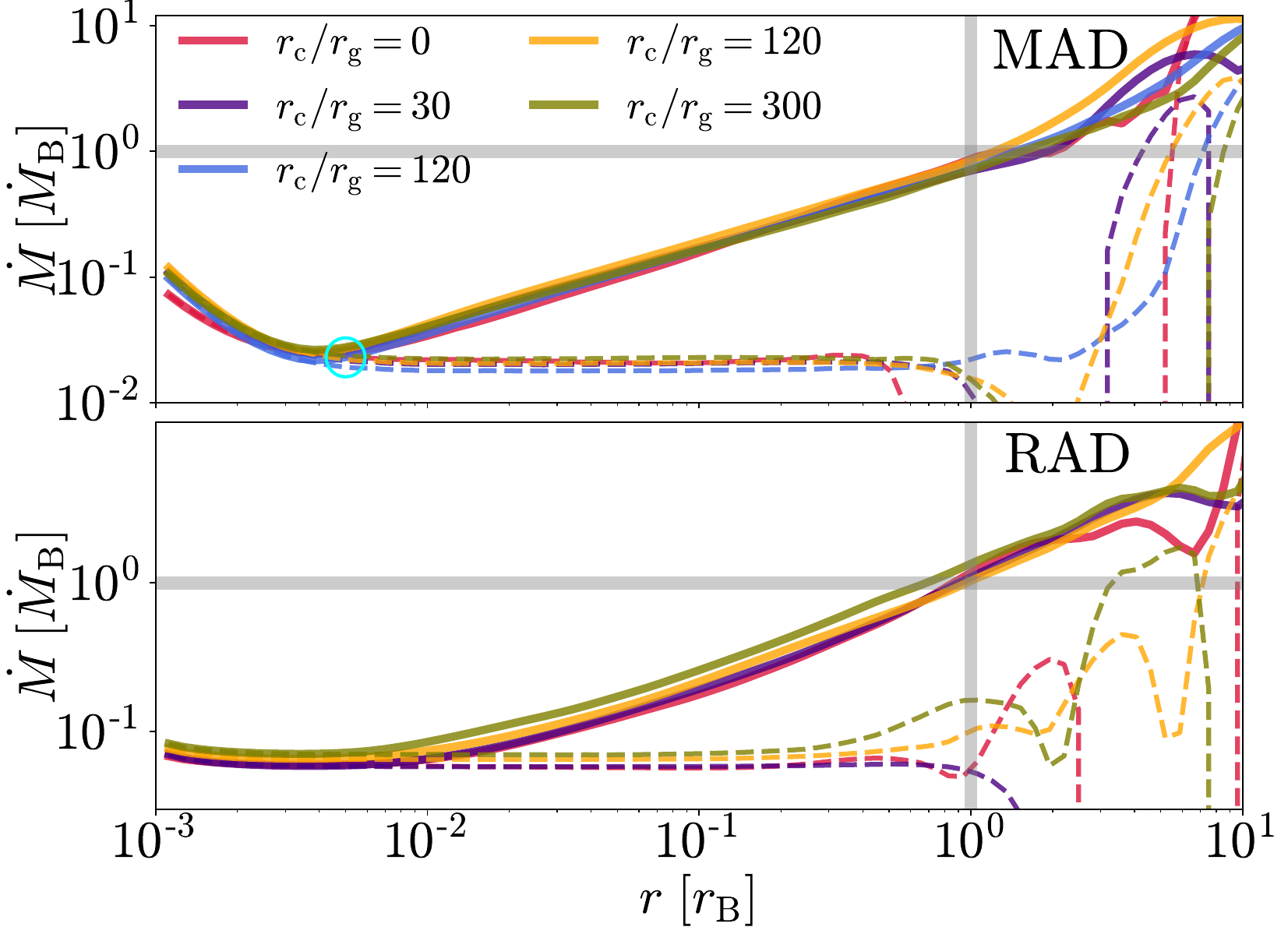}
\caption{The MAD and RAD power-law scalings are largely insensitive to the circularization radius, \rcirc, as seen for simulations with $\rb/\rg = 10^3$. The cyan circle marks where we measure \mdotbh, at $r=5\rg$, to avoid contamination by the density floors.}
\label{fig:all_rc}
\end{figure}

\begin{figure*}
\centering
\includegraphics[width=1\textwidth]{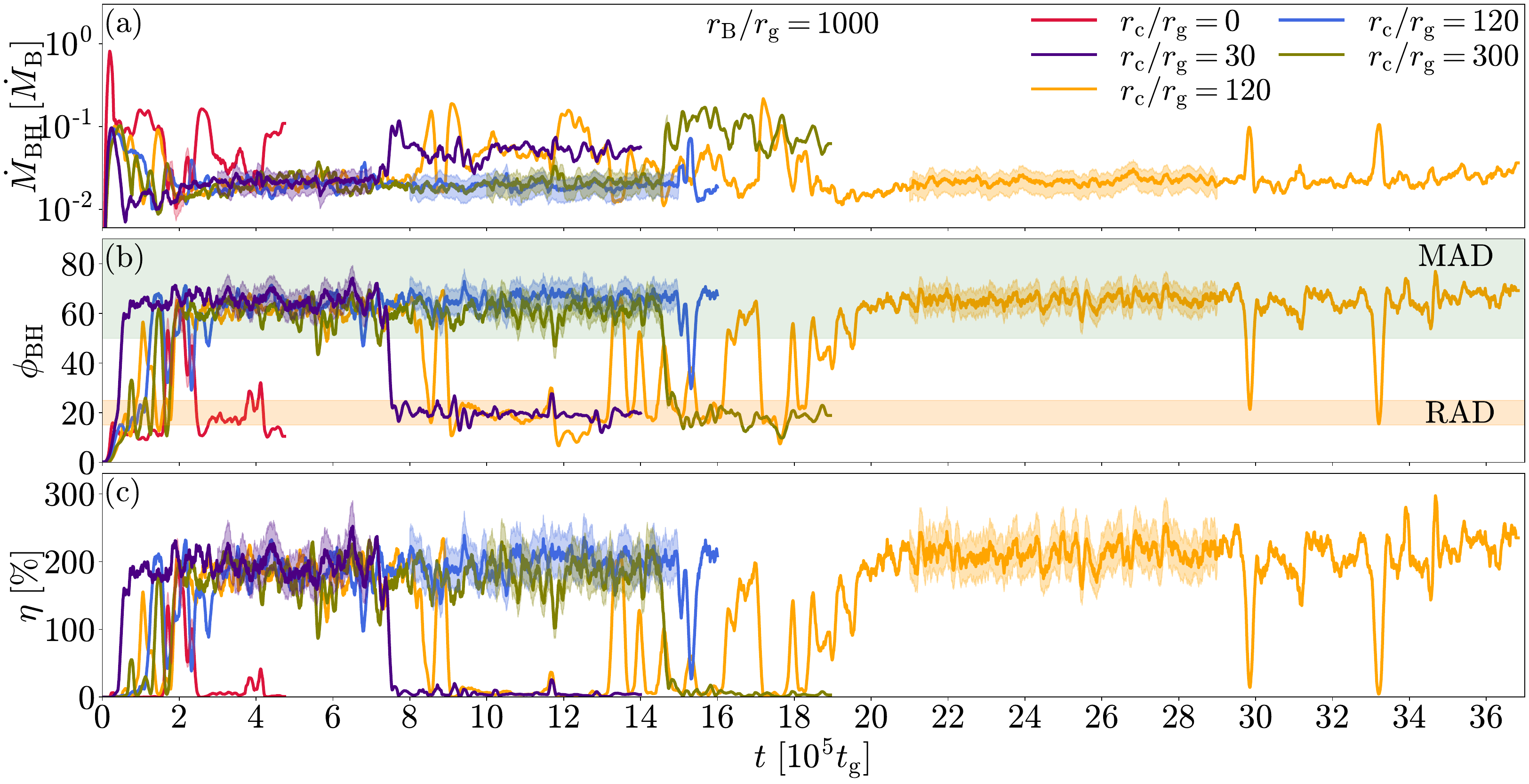}
\caption{ 
Changes in the circularization radius, $\rcirc/\rg$, have virtually no effect on the average fluxes in the MAD state. Namely, in the MAD state, all of our simulations, which sample a wide range of the circularization radius ($\rcirc/\rg=\{0,30,120,300\}$ for $\rb/\rg=1000$; see legend), display consistent values of BH mass accretion rate, $\mdotbh/\mdotb\simeq 0.02$ (a), BH magnetic flux, $\phibh \gtrsim 50$ (b), and BH outflow energy efficiency, $\eta \simeq 200\%$ (c).  
For $\rcirc = 0$ (red), the system enters MAD at $t \simeq 1.5 \times 10^5 \tg$ but transitions to RAD at $t \simeq 2.2 \times 10^5 \tg$, where $\mdotbh/\mdotb$ exceeds its MAD counterpart.  
For $\rcirc / \rg = 30$ (purple), MAD persists for $1 \times 10^5 \lesssim t / \tg \lesssim 7.5 \times 10^5$, before giving way to RAD. With $\rcirc / \rg = 120$ (blue), the system remains MAD until the end of the run. A repeated run, also with $\rcirc / \rg = 120$ but a different initial random seed (yellow), transitions to RAD at $t / \tg \simeq 9 \times 10^5$, and does not return back to the MAD state until $t \gtrsim 2 \times 10^6 \tg$. This RAD-to-MAD transition resembles the $\rb / \rg = 300$ case (Fig.~\ref{fig:mdot_vs_time}).  Our $\rcirc/\rg = 300$ model (dark green) remains in the MAD state at $2\times10^2\lesssim t/\tg \lesssim 1.5\times 10^6$, after which it transitions to the RAD state. For clarity, we smoothed all quantities over a timescale of $10^4\,\rg/c$ using a zeroth-order Savitzky–Golay filter.
}
\label{fig:mdot_vs_rcirc}
\end{figure*}

\section{Error estimation}\label{appendixc}
To estimate the error of the time-averaged values, we account for temporal correlations among data points. The standard error is typically the standard deviation, \( \sigma \), divided by the square root of the number of independent data points, \( N \). However, when data points are correlated, the effective number of independent samples decreases.

We quantify this reduction using the integrated autocorrelation time, \( \tau_{\text{auto}} \), normalized by the simulation sampling interval, \( \Delta t_{\rm sam} \):
\begin{equation}
\frac{\tau_{\text{auto}}}{\Delta t_{\rm sam}} = 1 + 2 \sum_{i=1}^{\infty} \rho(i\Delta t),
\end{equation}
where \( \rho(\tau) \) is the normalized autocorrelation function at time lag \( \tau \).

The error of the mean, incorporating the effect of autocorrelation, is then expressed as:

\begin{equation}
\sigma' = \sqrt{\frac{2 \tau_{\text{auto}}}{N}} \sigma
\end{equation}

We generally find a time lag, $\tau_{\text{auto}} \lesssim 50$, which increases the error by a factor of $\lesssim 10$, compared to the standard expression for the error of the mean (e.g., when treating all snapshots as independent). 


\bibliography{mybib}{}
\bibliographystyle{apsrev4-2}

\end{document}